\documentclass[11pt]{iopart}
\usepackage[utf8]{inputenc}
\usepackage[T1]{fontenc}		
\usepackage{comment}
\expandafter\let\csname equation*\endcsname=\relax 
\expandafter\let\csname endequation*\endcsname=\relax 
\usepackage{mathtools} 			
\usepackage{amssymb}
\usepackage{graphicx}

\usepackage{upgreek}
\usepackage{nicefrac}
\usepackage{braket}

\newcommand*{\dd}[1]{\mathrm{d}#1} 
\newcommand*{\defeq}{\coloneqq}

\DeclareMathOperator{\sinc}{sinc}
		
\newcommand*{\ind}{\chi}
\newcommand*{\ddelta}{\updelta}	

\newcommand*{\R}{\mathbb{R}}
\newcommand*{\C}{\mathbb{C}}

\newcommand*{\Z}{\mathbb{Z}}

\newcommand*{\iu}{\mathrm{i}\mkern1mu} 
\renewcommand{\Re}{\operatorname{Re}}

\newcommand*{\UU}{\mathrm{U}}

\DeclarePairedDelimiter\abs{\lvert}{\rvert}
\newcommand*{\matt}[1]{\begin{pmatrix} #1 \end{pmatrix}}

\usepackage[dvipsnames, x11names]{xcolor} 
\definecolor{Rosso}{cmyk}{0.3,1,1,0.2}	
\definecolor{Blu}{cmyk}{1,0.6,0,0.2}	
\definecolor{Verde}{cmyk}{1,0.21,1,0.2}

\usepackage{tikz}
\tikzset{>=latex}
\usetikzlibrary{calc, intersections, fadings, shadings,%
	positioning, decorations.pathreplacing, arrows, plotmarks, shapes.geometric,  shapes.geometric, arrows.meta, decorations.markings, bending, backgrounds}

\usepackage{pgfplots}
\pgfplotsset{/pgf/number  format/1000 sep={\,},%
		 compat=newest, trig format plots=rad, tick label style={font=\footnotesize},	label style={font=\small},}
\usepgfplotslibrary{colormaps, fillbetween, external, colorbrewer, groupplots}

\usepackage{hyperref}
\hypersetup{colorlinks=true, citecolor=Rosso, linkcolor=Blu}

\begin{document}

\title{Classical echoes  of quantum boundary conditions}

\author{Giuliano Angelone$^{1,2,*}$, Paolo Facchi$^{1,2,**}$, Marilena Ligab\`o$^{3,***}$}

\address{$^1$ Dipartimento di Fisica, Universit\`{a} degli Studi di Bari, I-70126 Bari, Italy}
\address{$^2$ INFN, Sezione di Bari, I-70126 Bari, Italy}
\address{$^3$  Dipartimento di Matematica, Universit\`{a}  degli Studi di Bari, I-70125 Bari, Italy}
\ead{$^*$giuliano.angelone@ba.infn.it, $^{**}$paolo.facchi@ba.infn.it, $^{***}$marilena.ligabo@uniba.it}

\begin{abstract}
We consider a non-relativistic particle in a one-dimensional box with all possible quantum boundary conditions that make the kinetic-energy operator self-adjoint. We determine the Wigner functions of the corresponding eigenfunctions and analyze in detail their classical limit in the high-energy regime. We show that the quantum boundary conditions split into two classes: all local and regular boundary conditions collapse to the same classical boundary condition, while singular non-local boundary conditions slightly persist in the classical limit.
\end{abstract}


\maketitle

\section{Introduction}
The phase-space formulation of quantum mechanics allows to represent states and operators as  functions on classical phase-space
\cite{Wi32}. It has various applications, ranging from quantum optics, quantum chaos and quantum computing to classical optics and signal analysis  \cite{HiCoSc84, Lee95}. In quantum physics, phase-space methods have also been used to characterize the non-classicality of quantum states, to identify and reconstruct states via quantum tomography, and to understand the quantum-to-classical transition and the correspondence principle \cite{CaKi87, HaKo89, MMSSV05, KoZeGl20}.

Between many possible quantum (quasi-)probability distributions, the Wigner function arguably gives the most natural phase-space representation of quantum mechanics. In spite of a long history of research, the theory of Wigner functions for systems on a manifold (or phase-space) with non-trivial topology, as well as having boundaries, is still not complete. For example, group-theoretical approaches have recently been applied for the Wigner function on the cylinder $\mathbb{S}^1\times \R$ \cite{PrBrTo14, KoLa21}, the discrete cylinder $\Z\times\R$ \cite{ZhVo03, RSSK11} and the torus $\mathbb{S}^1\times \mathbb{S}^1$  \cite{Li16}, whereas the deformation quantization approach is usually employed for manifold with boundaries, see e.g.~\cite{DiPr02, KrWa05, DiPr07, DiPoPr11}.

In this paper, we are interested in two related subjects: (i) the study of the Wigner function for eigenfunctions of the kinetic-energy operator (i.e.\ the free Hamiltonian) acting in a one-dimensional box with general self-adjoint boundary conditions, and (ii) the analysis of these Wigner functions in the classical limit, that is in the high-energy regime. Preliminary results in this direction have already been obtained: the Wigner function has been studied in \cite{CaKrMa91, BeDoRo04} for the one-dimensional box with Dirichlet boundary conditions, and in \cite{Wal07, AlWi21} for the half-line with Robin boundary conditions. Besides,  in \cite{Rob95, Rob02, BeMaGa13} the classical limit of the position and momentum probability distributions for the one-dimensional box with Dirichlet boundary conditions have been investigated.

The paper is organized as follows. After introducing in Sec.~\ref{sec:box} the free Hamiltonian with general self-adjoint boundary conditions, in Sec.~\ref{sec:wigner} we explicitly compute the Wigner functions associated with the eigenfunctions of the Hamiltonian, and describe how to determine their classical limit. In  Sec.~\ref{sec:asymptotic}, then, we classify the possible classical limits by analyzing the asymptotic properties of the spectrum in the high-energy regime.
Finally, in  Sec.~\ref{sec:discussion} we discuss the results and compare the classical limits with corresponding classical probability distributions.

\section{A quantum particle in a box}\label{sec:box}
We consider a quantum particle of mass $m$, confined in a one-dimensional box of unit length, 
namely the interval $J=[-1/2,1/2]$.  This system is formally described by the kinetic-energy operator,  $x=\tilde{x}/L$) 
\begin{equation}\label{eq:H}
H= -\frac{\hbar^2}{2m}\frac{\mathrm{d}^2}{\mathrm{d}x^2}\,,
\end{equation}
which acts on a proper subspace of the Hilbert space $L^2(J)$. 
As it is well-known, see e.g.~\cite{Te14, BFV01}, Eq.~\eqref{eq:H} prescribes the action of $H$ only in the \emph{bulk} of the system. The Hamiltonian $H$ should indeed be equipped with suitable boundary conditions (BCs), specifying the behavior of the particle at the boundary of the interval, in order to generate a well-defined quantum dynamics. In quantum mechanics the possible  BCs, encoded in the domain $\mathcal{D}(H)$ of $H$, cannot be arbitrary, but are constrained by the requirement that $H$ must be a \emph{self-adjoint} operator, i.e.\ $\mathcal{D}(H)=\mathcal{D}(H^*)$ and $H=H^*$. Indeed, self-adjointness is a necessary and sufficient condition for a (Hermitian) operator to have a purely real spectrum and to generate a unitary dynamics. 

Different domains correspond to different behaviors of the particle at the boundary, give rise to different dynamics and represent different physical situations. All the self-adjoint realizations  of the operator~\eqref{eq:H} are known to be in one-to-one correspondence with the set of $2\times 2$ unitary matrices $U\in\UU(2)$~ \cite{AIM05, AIM15, IBP15, FGL18b}. Each of these realizations, which we henceforth denote by $H_U$, is defined on the domain
\begin{equation}\label{eq:domain}
\mathcal{D}(H_U)=\{\psi \in \mathcal{H}^2(J) : \Psi_-=U\Psi_+\}\,,\qquad U\in\UU(2)\,,
\end{equation}
where $\mathcal{H}^2(J)$ is the space of wave functions $\psi$ with square-integrable first and second derivative, $\psi'$ and $\psi''$, on the interval $J=[-1/2,1/2]$, whereas 
\begin{equation}\label{eq:bc}
\Psi_-=U\Psi_+
\end{equation}
is defined in terms of the following boundary values\footnote{In some related works, as e.g.~\cite{AIM05, AIM15, IBP15}, the alternative parametrization $\Psi_+=-\tilde{U}\Psi_-$ is adopted, with a given $\tilde{U}\in\UU(2)$; here, following \cite{FGL18b, FGL18}, we find convenient to put $\tilde{U}=-U^{\dagger}$.}
\begin{align}\label{eq:Psipm}
\Psi_\pm \defeq
\begin{pmatrix}
-\psi'(-\frac{1}{2})\pm \iu  \psi(-\frac{1}{2})\\[3pt] 
+\psi'(\frac{1}{2})\pm \iu  \psi(\frac{1}{2})
\end{pmatrix}\,. 
\end{align}

\begin{figure}[tbp]
\centering
\includegraphics{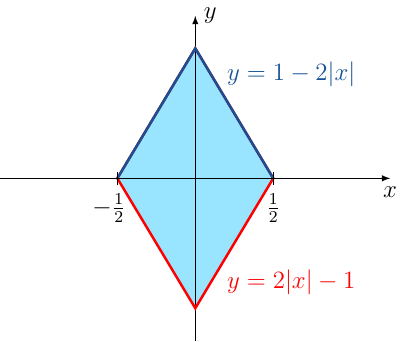}
\caption{Integration region of Eq.~\eqref{eq:W} (shaded area): as $x$ varies in $J=[-1/2,1/2]$ the integrand function contributes to the integral only  for 
$-1/2\leq x+y/2 \leq 1/2$ and $-1/2\leq x-y/2 \leq 1/2$, that is for $2\abs{x}-1\leq y \leq 1-2\abs{x}$.}
\label{fig:intdomain}
\end{figure}

Let us now introduce the Wigner function associated with a wave function $\psi\in L^2(\R)$, \cite{Wi32,HiCoSc84, Lee95, Cas08}. The Wigner function $W\psi$ represents the joint quasi-probability distribution of position and momentum in the state $\psi$ and it is given by
\begin{equation}
W\psi(x,p)
\defeq \frac{1}{2\pi \hbar}\int_{\R}
\e^{-\iu p y /\hbar }
\psi\bigl(x+\tfrac{y}{2}\bigr)
\overline{\psi\bigl(x-\tfrac{y}{2}\bigr)}
\,\dd{y}
\end{equation}
where $(x,p)\in\R^2$. For a wave function $\psi$ spatially confined in the interval $J$, i.e.\ for an element of $L^2(J)$, the associated Wigner function of $\psi$ can be computed considering the function defined on $\R$ that coincides with $\psi$ in the interval $J$ and vanishes outside. With this procedure one obtains
\begin{subequations}\label{eq:W} 
\begin{align}
W\psi(x,p)
&= \frac{1}{2\pi \hbar}\int_{\R}
\e^{-\iu p y /\hbar }
\psi\bigl(x+\tfrac{y}{2}\bigr)
\overline{\psi\bigl(x-\tfrac{y}{2}\bigr)}
\,\dd{y}
\\
&=\frac{\ind(x)}{2\pi \hbar}\int_{2\abs{x}-1}^{1-2\abs{x}}
\e^{-\iu p y/\hbar}
\psi\bigl(x+\tfrac{y}{2}\bigr)
\overline{\psi\bigl(x-\tfrac{y}{2}\bigr)}
\,\dd{y}\,,
\end{align}
\end{subequations}
where $\ind$ is the characteristic function of the interval $J=[-1/2,1/2]$, i.e.\ $\ind(x)=1$ if $|x|\leq 1/2$ and $\ind(x)=0$ if $|x|> 1/2$, see Fig.~\ref{fig:intdomain}.
We stress that, although being defined for $(x,p)\in\R^2$, by construction the above Wigner function vanishes for $\abs{x}>1/2$, i.e.\ outside of the box.\footnote{Interestingly, Eq.~\eqref{eq:W} can  also be obtained by applying a ``regularization'' procedure: in  \cite{BW10}, e.g., a particle moving freely on the half-line is treated as moving on the full line in the presence of an infinite potential wall, the latter being realized as a limit of a smooth (Morse) potential.}

In the following we are interested both in the explicit expression of the Wigner function for an eigenfunction of $H_U$, which will be the main topic of Sec.~\ref{sec:wigner}, and in its behavior in the \emph{classical limit}, which we will discuss in Sec.~\ref{sec:asymptotic}. 
Before moving on, however, we spend a few words on the allowed quantum BCs, giving some examples in Sec.~\ref{sec:genbc} and introducing a useful parametrization of $\UU(2)$ in Sec.~\ref{sec:param}.

\subsection{Topology of quantum boundary conditions}\label{sec:genbc}
It is convenient to rewrite the boundary values in~\eqref{eq:Psipm} as
\begin{align}
\Psi_\pm=\Psi'\pm\iu \Psi\,,&&
\Psi \defeq \begin{pmatrix}
\psi(-\frac{1}{2})\\[3pt] \psi(\frac{1}{2})
\end{pmatrix}\,,&&
\Psi' \defeq \begin{pmatrix}
- \psi'(-\frac{1}{2})\\[3pt] \psi'(\frac{1}{2})
\end{pmatrix}\,,&&
\end{align}
so that, if the matrix $I-U$ is invertible, the BC in Eq.~\eqref{eq:bc} can also be  expressed  as
\begin{align}\label{invCaylay}
\Psi'=M_U\Psi\,, \qquad M_U=\iu (I+U)(I-U)^{-1}\,,
\end{align}
where $M_U$, being the inverse Cayley transform of $U$, is an Hermitian matrix. For a more general inversion formula, holding also when $I-U$ is not invertible, see e.g.\ Eq.~(19) of~\cite{FGL18}. Two interesting families of BCs are given respectively by (symmetric) Robin conditions
\begin{align}\label{eq:Robinbc}
U_\textup{R}(\alpha)\defeq\e^{\iu \alpha}I\,,\qquad \Psi'=-\cot\Bigl(\frac{\alpha}{2}\Bigr)\Psi\,,
\end{align}
with $\alpha \in {[0,2\pi)}$, that reduce to Dirichlet ($\Psi=0$) and Neumann ($\Psi'=0$) BCs respectively for $\alpha=0$ and $\alpha=\pi$, and by pseudo-periodic BCs
\begin{align}\label{eq:pseudobc}
U_{\textup{pp}}(\alpha)\defeq -\begin{pmatrix}
0 & \e^{-\iu\alpha} \\  \e^{\iu\alpha} &0
\end{pmatrix}\,, \qquad \begin{cases}
\psi(\frac{1}{2})=\e^{\iu\alpha}\psi(-\frac{1}{2})\\[3pt]
\psi'(\frac{1}{2})=\e^{\iu\alpha}\psi'(-\frac{1}{2})
\end{cases},
\end{align}
with $\alpha \in {[0,2\pi)}$, that in turn reduce to periodic and anti-periodic conditions when $\alpha=0$ and $\alpha=\pi$, respectively. 

\begin{figure}[tbp]
\centering
\includegraphics{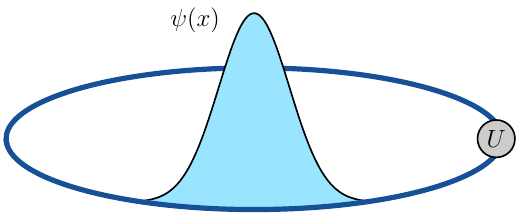}
\caption{A quantum particle in a ring with a junction.}
\label{fig:ring}
\end{figure}

As the reader may have noticed, BCs can be either local or non-local: Robin BCs, e.g., are local, as they do not mix the boundary values of $\psi$ at the left edge $x=-1/2$ with those at the right edge $x=1/2$, whereas pseudo-periodic BCs are non-local. Differently from local BCs, which physically model a particle in a box, non-local BCs are actually related to the physics of a particle in a ring.  Arbitrary BCs, thus, can be realized in a ring with a junction, the matrix $U$ encoding the physical properties of the latter, see Fig.~\ref{fig:ring}. Note that this setup describes also local BCs, as the junction may eventually act as an impenetrable barrier that ``decouples'' the left edge from the right one.

More generally, let us stress that BCs are often crucial to determine the spatial topology of a quantum system, see e.g.~\cite{BaBiMa95,AIM05, ShWi12, AFMP13,IbLlPe15,FGMSS18,movingwalls,bangalectures} for further details.

\subsection{Parametrization of the quantum boundary conditions}\label{sec:param}
All the unitary matrices in $\UU(2)$ can be parametrized by using five real parameters \cite{isoboundary}:
\begin{align}\label{eq:Upar}
\e^{\iu\eta}\begin{pmatrix}
m_0+\iu m_3 & m_2+\iu m_1 \\ -m_2+\iu m_1 & m_0-\iu m_3
\end{pmatrix}
\,, 
\end{align}
with $\eta\in[0,2\pi)$, and  $m_0, m_1, m_2, m_3 \in \mathbb{R}$ such that
\begin{equation}\label{eq:mconstraint}
m_0^2+m_1^2+m_2^2+m_3^2=1\,.
\end{equation}
In order to obtain a one-to-one parametrization of $\UU(2)$, the values $(\eta,m_0,m_1, m_2, m_3)$ and $(\eta+\pi, -m_0, -m_1, -m_2, -m_3)$ have to be identified, as they give the same matrix:
\begin{equation}\label{eq:ident}
\e^{\iu\eta}\begin{pmatrix}
m_0+\iu m_3 & m_2+\iu m_1 \\ -m_2+\iu m_1 & m_0-\iu m_3
\end{pmatrix}
=\e^{\iu(\eta+\pi)}\begin{pmatrix}
-m_0-\iu m_3 & -m_2-\iu m_1 \\ m_2-\iu m_1 & -m_0+\iu m_3
\end{pmatrix}\,.
\end{equation}
To achieve this, we henceforth restrict $\eta\in\left[0,\pi\right)$.  On the other hand, Eq.~\eqref{eq:mconstraint} tells us that only four parameters are actually independent and, since the pair $(m_0,m_1)$ always takes values in the unit disk $D=\{(x,y) \in \mathbb{R}^2: x^2+y^2\le 1\}$,
it is convenient to express $m_2$ and $m_3$ in terms of a new parameter $\beta\in\left[0,2\pi\right)$:
\begin{align}\label{eq:beta}
m_2=\sqrt{1-m_0^2-m_1^2}\,\cos(\beta)\,, \qquad  m_3=\sqrt{1-m_0^2-m_1^2}\, \sin(\beta)\,.
\end{align}
Therefore we have that
\begin{equation}
\UU(2)=\left\{U(\eta, m_0,m_1,\beta) \,:\,  \eta \in [0,\pi),\, (m_0,m_1) \in D,\, \beta \in [0,2\pi) \right\},
\end{equation}
where for all $\eta \in [0,\pi)$, $(m_0,m_1) \in D $ and $ \beta \in [0,2\pi)$:
\begin{equation}\label{eq:param}
U(\eta, m_0,m_1,\beta)\defeq \e^{\iu\eta}\begin{pmatrix}
m_0+\iu \sqrt{1-m_0^2-m_1^2}\,\sin(\beta) & \sqrt{1-m_0^2-m_1^2}\,\cos(\beta)+\iu m_1 \\ -\sqrt{1-m_0^2-m_1^2}\,\cos(\beta)+\iu m_1 & m_0-\iu \sqrt{1-m_0^2-m_1^2}\, \sin(\beta)
\end{pmatrix}\,.
\end{equation}
Notice that if $m_0^2+m_1^2=1$ then the matrix $U(\eta, m_0,m_1,\beta)$ does not depend on $\beta$, in that case we will fix $\beta=0$.

\begin{figure}[tp]
\centering
\includegraphics{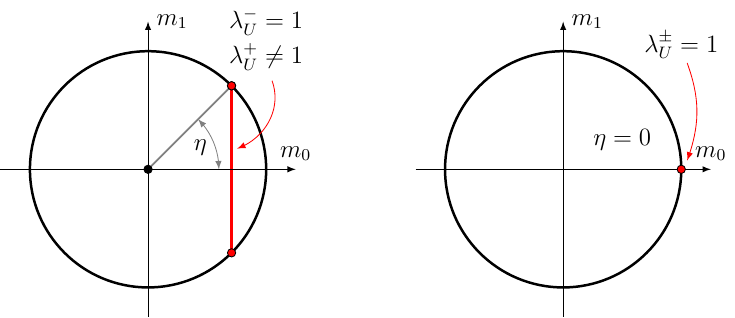}

\caption{Boundary conditions $U(\eta, m_0,m_1,\beta)$ having at least one eigenvalue $\lambda_U^\pm$ equal to $1$ are shown, in red, in the parameter space $(m_0,m_1)\in D$, for $0<\eta<\pi$ (left) and $\eta=0$ (right).}
\label{fig:cayley}
\end{figure}

Observe that Robin BCs~\eqref{eq:Robinbc} correspond to
\begin{equation}\label{eq:RobinbcU}
U_\textup{R}(\alpha)=\begin{cases}
U(\alpha, 1, 0,0)\,, &\text{if }0\le \alpha<\pi\,,\\
U(\alpha-\pi, -1, 0,0)\,, &\text{if }\pi\le \alpha< 2\pi\,,
\end{cases}
\end{equation}
whereas pseudo-periodic BCs~\eqref{eq:pseudobc} correspond to
\begin{equation}\label{eq:pseudobcU}
U_{\textup{pp}}(\alpha)=\begin{cases}
U\bigl(\tfrac{\pi}{2}, 0, \cos(\alpha),0\bigr)\,, &\text{if }0\le \alpha<\pi\,,\\
U\bigl(\tfrac{\pi}{2}, 0, \cos(\alpha),\pi\bigr)\,, &\text{if }\pi\le \alpha< 2\pi\,.
\end{cases}
\end{equation}
The eigenvalues of $U(\eta, m_0,m_1,\beta)$ depend only on $\eta $ and $m_0$ and are given by
\begin{align}
\lambda_U^\pm&\defeq\exp{\iu \left[\eta\pm\arccos(m_0)\right]}\,, 
\end{align}
and in particular
\begin{subequations}
\begin{align}
\lambda_U^-=1 &\quad\Leftrightarrow\quad
m_0=\cos (\eta) 
\,,\label{eq:ceig1}\\ 
\lambda_U^+=1 &\quad\Leftrightarrow\quad \eta=-\arccos(m_0)=0
\quad\Leftrightarrow\quad 
\eta=0\,,\quad m_0=1 
\,.\label{eq:ceig3}
\end{align}
\end{subequations}
The above values are relevant since the inverse Cayley transform of $U(\eta, m_0,m_1,\beta)$ is singular whenever $I-U(\eta, m_0,m_1,\beta)$ is not invertible, that is when $\lambda_U^-$ or $\lambda_U^+$ is equal to $1$. Otherwise, it is the well defined Hermitian matrix
\begin{equation}\label{eq:cayley}
M_{U(\eta,m_0,m_1,\beta)}=\frac{1}{m_0-\cos\eta}\begin{pmatrix}
-\sin(\eta)+r\sin(\beta) & m_1-\iu r\cos(\beta) \\ m_1+\iu r\cos(\beta) & -\sin(\eta)-r\sin(\beta)
\end{pmatrix}\,,
\end{equation}
where $r=\sqrt{1-m_0^2-m_1^2}$.
In Fig.~\ref{fig:cayley}, BCs having at least one eigenvalue equal to $1$ are  represented in the parameter space of $(m_0,m_1)$, for a given value of $\eta$. Note that, in particular, the Dirichlet condition $U_\textup{R}(0)=I$ is the only one having $\lambda_U^-=\lambda_U^+=1$. 

For later convenience, we introduce the definition of \emph{regular} and \emph{singular} BCs.
We say that the unitary matrix $U(\eta, m_0,m_1,\beta)$ (and the corresponding BC) is \emph{singular} if the eigenvalues satisfy $\lambda_U^-=1$ and $\lambda_U^+\neq 1$, that is if
\begin{equation}\label{eq:sing}
m_0= \cos(\eta)\neq 1\,,
\end{equation}
while it is \emph{regular} otherwise. In other words the matrices in $\UU(2)$ having zero or two eigenvalues equal to $1$ are regular, while the ones having just one eigenvalue $1$ are singular. 
Besides, excluding the case of the identity matrix $I$ (corresponding to the Dirichlet condition), which is regular by definition but does not admit the inverse Cayley transform, a generic unitary matrix $U$ is regular if and only if it admits the inverse Cayley transform. 
Notice that the matrices corresponding to Robin BCs in Eq.~\eqref{eq:RobinbcU} are regular (including the Dirichlet condition), while the matrices corresponding to pseudo-periodic BCs in Eq.~\eqref{eq:pseudobc} are singular.

\section{Wigner functions}\label{sec:wigner}
In this section we explicitly determine the Wigner function of an eigenfunction of $H_U$, with $U=U(\eta, m_0,m_1, \beta) \in \UU(2)$, $\eta \in [0,\pi), (m_0,m_1) \in D$ and $\beta \in [0,2\pi)$,  that is a non-zero solution of the eigenvalue equation 
\begin{equation}\label{eq:eig}
H_U \psi=E\psi.
\end{equation}
More in detail, after solving in Sec.~\ref{sec:spectrum}  the spectral problem of $H_U$, by determining its eigenfunctions in terms of the zeroes of a certain spectral function, in Sec.~\ref{sec:classlim} we compute the corresponding Wigner functions, and analyze some of their properties in the high-energy regime. Then, in Sec.~\ref{sec:classical} we review the phase-space description of a classical particle in a box in order to compare the results.
 
\subsection{Spectral problem}\label{sec:spectrum}
The eigenvalue equation~\eqref{eq:eig} can be rewritten as the ordinary differential equation
\begin{equation}\label{eq:diffeq}
\psi''+\epsilon \psi=0\,,\qquad \epsilon= 2mE/\hbar^2\in\R\,,
\end{equation}
further supplied by the BC in Eq.~\eqref{eq:bc}. Here, $\epsilon$ represents the dimensionless energy. For $\epsilon\neq0$, the eigenvalue equation~\eqref{eq:diffeq} has a general solution of the form
\begin{align}\label{eq:eigenfunction}
\psi_U(x; \epsilon)=\frac{1}{N_U(k)} \left(C^+_U(k)\e^{\iu  k x}+C^-_U(k)\e^{-\iu  k x}\right)\,,  \qquad x \in (-\tfrac{1}{2},\tfrac{1}{2}) \,,
\end{align}
where $C^\pm_U(k)\in\C$, $N_U(k)\in \R$ is a normalization constant, and
\begin{equation}\label{eq:keps}
 k\defeq  \e^{\iu\arg(\epsilon)/2}\sqrt{\abs{\epsilon}}
\end{equation}
is the dimensionless wave number. Differently from $\epsilon$ which is always real, $ k$ can be either real or purely imaginary, respectively when $\epsilon\ge 0$ or $\epsilon<0$. We recall that, in general,  the eigenvalues of $H_U$ accumulate to $+\infty$ and can be at most doubly degenerate (see e.g.\ Theorem~10.6.1 of~\cite{Ze05}), thus depending on $U$ there can be at most two vanishing eigenvalues $\epsilon=0$. Moreover, the sum of the multiplicities of the negative eigenvalues is at most two~\cite{BFV01,isoboundary}. 

\begin{figure}
\centering
\includegraphics[width=\linewidth]{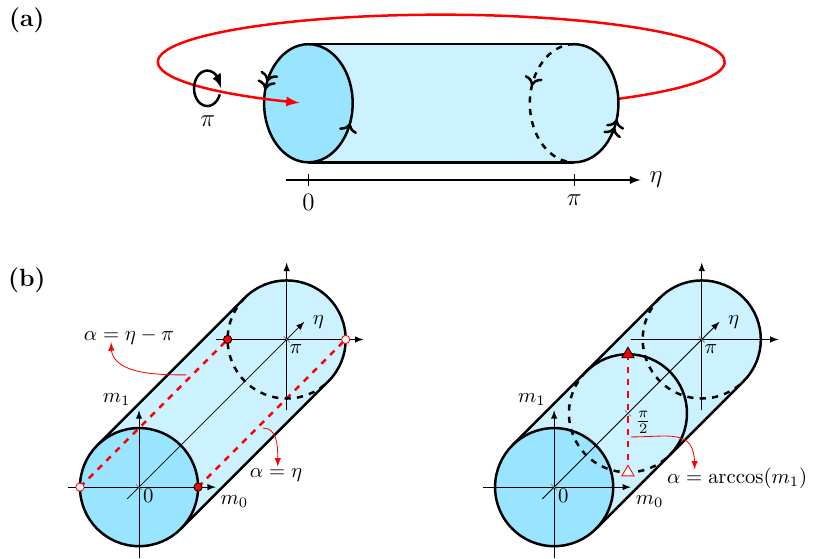}
\caption{(a) The spectral space $\Sigma$ can be constructed by gluing the two bases of the solid cylinder $[0,\pi]\times D$ after applying a global twist of angle $\pi$.  The twist emerges as a consequence of the identification described by Eq.~\eqref{eq:ident}. (b) Some example of BCs depicted in the parameter space $[0,\pi]\times D$.  Left: Robin conditions $\Psi'=-\cot(\frac{\alpha}{2})\Psi$ (red dashed lines), Neumann condition (\protect\tikz\protect\draw[red, fill=white] circle (0.1);) and Dirichlet condition (\protect\tikz\protect\draw[black, fill=red] circle (0.1);), see Eq.~\eqref{eq:RobinbcU}. Right: pseudo-periodic conditions $\psi(\frac{1}{2})=\e^{\iu\alpha}\psi(-\frac{1}{2})$ (red dashed line), periodic condition (\protect\tikz\protect\node[draw=black, fill=red, regular polygon, regular polygon sides=3,inner sep=1.5pt] at (1.25,0.25) {};) and anti-periodic condition (\protect\tikz\protect\node[draw=red, fill=white, regular polygon, regular polygon sides=3,inner sep=1.4pt] at (1.25,0.25) {};), see Eq.~\eqref{eq:pseudobcU}.}
\label{fig:sigma}
\end{figure}

To impose the BC in Eq.~\eqref{eq:bc}, after substituting Eq.~\eqref{eq:eigenfunction} in the expression~\eqref{eq:Psipm} of the boundary values $\Psi_\pm$, we rewrite the latter as
\begin{align}\label{eq:Apm}
\Psi_\pm=\frac{1}{N_U}A_\pm(\epsilon)\matt{C^+_U(k)\\ C^-_U(k)}\,,&& 
A_\pm(\epsilon)\defeq \pm \iu \matt{
(1\mp  k)\e^{-\iu  k/2} & (1\pm  k)\e^{\iu  k/2} \\ 
(1\pm  k)\e^{\iu  k/2} & (1\mp  k)\e^{-\iu  k/2}
}\,.
\end{align}
The BC $\Psi_-=U\Psi_+$ can then be expressed as the homogeneous system
\begin{equation}\label{eq:AUA}
\left[A_-(\epsilon)-UA_+(\epsilon)\right]\matt{C^+_U(k)\\ C^-_U(k)}=0\,,
\end{equation}
whose non-trivial solutions are obtained by requiring that
\begin{equation}
F_U(\epsilon)\defeq\det\bigl(A_-(\epsilon)-UA_+(\epsilon)\bigr)=0\,.
\end{equation}
In other words, the non-vanishing eigenvalues of $H_U$ correspond to the real zeroes of the \emph{spectral function} $F_U(\epsilon)$.\footnote{We mention that the definition of the spectral function can be modified to account also for the zero eigenvalues, see  \cite{isoboundary} for details.} In terms of the parametrization~\eqref{eq:param}, it is known \cite{BFV01,AIM15, isoboundary} that, for $\epsilon\neq0$, 
\begin{equation}\label{eq:FU}
F_U(\epsilon) = \sin (k)\bigl[ k^2\bigl(\cos(\eta)-m_0\bigr)+\cos(\eta)+m_0\bigr]-2 k \bigl[m_1-\sin(\eta)\cos (k)\bigr]\,,
\end{equation}
with $ k=\e^{\iu\arg(\epsilon)/2}\sqrt{\abs{\epsilon}}$ as in Eq.~\eqref{eq:keps}. 
As it turns out, the spectrum $\sigma(H_U)$ depends only on three of the four independent parameters characterizing the matrix $U\in\UU(2)$. Namely, it depends on $\eta$, $m_0$ and $m_1$, which by now we call \emph{spectral parameters}, but not on $\beta$, the \emph{non-spectral parameter}.  Seen as a manifold, the \emph{spectral space}
\begin{equation}
\Sigma\defeq [0,\pi] \times D
\end{equation}
has the same structure of a (twisted) solid torus \cite{isoboundary}, see  Fig.~\ref{fig:sigma}~(a).  In Fig.~\ref{fig:sigma}~(b), as an example, we represent in $\Sigma$ both Robin and pseudo-periodic conditions.

 The zeroes of $F_U(\epsilon)$ can be found analytically only for some particular BCs. Nevertheless, as we will show, their asymptotic behavior in the high-energy regime $\epsilon\to+\infty$ follows a simple pattern. Since we are interested in the classical limit, that is in the high-energy regime, from now on we will focus only on the positive part of the spectrum:
\begin{equation} 
\sigma_+(H_U)=\{E\in \sigma(H_U):E>0\}\,.
\end{equation}
 For the time being, let us denote with 
\begin{equation}\label{eq:epsn}
\bigl(\epsilon_n(\eta, m_0,m_1)\bigr)_{n \geq 1}
\end{equation}
the sequence of the positive zeroes of $F_U(\epsilon)$ and with
\begin{equation}\label{eq:kn}
 k_n(\eta, m_0,m_1)\defeq\sqrt{\epsilon(\eta, m_0,m_1)},  \quad n \geq 1\,,
\end{equation}
the corresponding wave numbers, so that we have
\begin{equation}
\sigma_+(H_U)=\biggl\{\frac{\hbar^2}{2m} k^2_n(\eta, m_0,m_1) :  n \geq 1 \biggr\}\,.
\end{equation}

To fix the expression of the eigenfunctions, we need to  explicitly determine  the coefficients $C^\pm_U(k)$ and the normalization $N_U(k)$. By using Eq.~\eqref{eq:AUA} we find
\begin{align}
C^\pm_{U}(k)&=\pm\e^{\pm\iu\frac{ k}{2}}\bigl[(1\pm  k)(m_0+\iu m_3) +(1\mp  k) (\e^{-\iu\eta}+\e^{\mp\iu  k }(m_2+\iu m_1)) \bigr]\,, \label{eq:Cpm}\\
N_U^2(k)&= \int_{-\frac{1}{2}}^{\frac{1}{2}} \abs{C_U^+(k) \e^{\iu  k x}+C_U^-(k)\e^{-\iu  k x}}^2\,\dd{x} \nonumber \\
&=\abs{C_U^+(k)}^2+\abs{C_U^-(k)}^2+2\frac{\sin( k)}{ k}\Re\Bigl(C_U^+(k)\overline{C_U^-(k)}\Bigr)\,,\label{eq:N}
\end{align}
where $m_2=\sqrt{1-m_0^2-m_1^2}\cos (\beta)$ and $m_3=\sqrt{1-m_0^2-m_1^2}\sin (\beta)$ as in Eq.~\eqref{eq:beta}. The above expressions reveal that, differently from the spectrum, the coefficients of the eigenfunctions in Eq.~\eqref{eq:eigenfunction} \emph{do} actually depend on the non-spectral parameter $\beta$. In conclusion, the function
 \begin{align}\label{eigenfnU}
 \psi_{U,n}(x)&\defeq\psi_U(x; \epsilon_n(\eta,m_0,m_1))\notag\\ 
 &=\frac{1}{N_{U,n}} \left(C_{U,n}^+ \e^{\iu  k_n(\eta,m_0,m_1) x}+C^-_{U,n}\e^{-\iu  k_n(\eta,m_0,m_1) x}\right)\,, 
 \end{align}
where $C_{U,n}^\pm\defeq C_{U}^\pm(k_n(\eta,m_0,m_1))$ and $N_{U,n}\defeq N_U(k_n(\eta,m_0,m_1))$,  is a normalized eigenfunction of $H_U$ corresponding to the eigenvalue $\hbar^2 \epsilon_n(\eta,m_0,m_1)/(2 m)$. For later convenience let us observe that, in the expression~\eqref{eq:N} for the normalization $N_{U,n}$, the interference term proportional to  $\sin( k_n(\eta,m_0,m_1))/ k_n(\eta,m_0,m_1)$ is negligible for large $n$, thus we have
\begin{equation}\label{eq:normC}
\lim_{n \to +\infty} \frac{\abs{C_{U,n}^+}^2+\abs{C_{U,n}^-}^2}{N_{U,n}^2}=1\,.
\end{equation}

\subsection{Classical limit of the Wigner functions}\label{sec:classlim} 
\begin{figure}[tbp]
\centering
\includegraphics{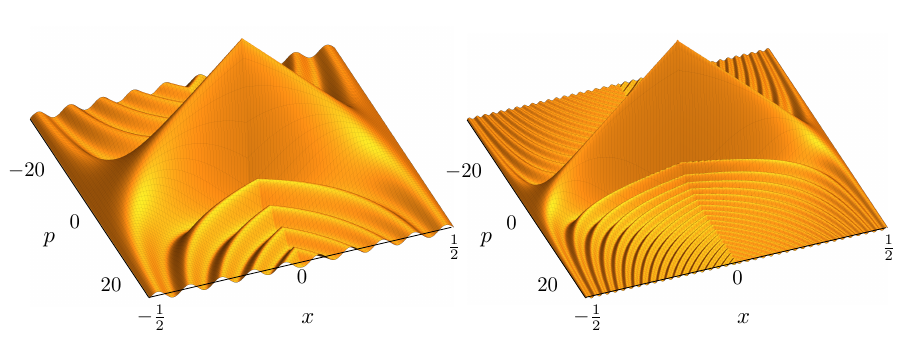}

\caption{Plot of $\hbar f_{0,n}(x,p)$ for $\hbar=1$ (left) and $\hbar=1/4$ (right).}
    \label{fig:fwign}
\end{figure}

At this point we are ready to compute the Wigner function associated with an eigenfunction~\eqref{eigenfnU}. Using the definition in Eq.~\eqref{eq:W} we obtain
\begin{equation}\label{eq:Wn}
W \psi_{U,n}=\frac{1}{N_{U,n}^2}\Bigl[ 
\abs{C^{+}_{U,n}}^2 f_{1,n} +
\abs{C^{-}_{U,n}}^2 f_{-1,n} +
2\Re\Bigl(C^{+}_{U,n}\overline{C^{-}_{U,n}}\e^{2\iu  k_n(\eta, m_0,m_1) x}\Bigr) f_{0,n}\Bigr] \,,
\end{equation}
where for each $s\in\{-1,0,1\}$:
\begin{equation}\label{eq:fn2}
f_{s,n}(x,p)\defeq\frac{\triangle(x)}{\pi\hbar}
\sinc\biggl(\frac{1}{\hbar}(p-s \hbar k_n)(1-2\abs{x})\biggr)\,,  \quad \forall x,p \in \mathbb{R},
\end{equation}
with the \emph{triangular envelope} and the \emph{sinc} functions being respectively given by
\begin{align}
\triangle(y)\defeq \ind(y)(1-2\abs{y})\,,&&\sinc(y)\defeq \frac{\sin y}{y},\, && \forall y \in \mathbb{R}.
\end{align}
A plot of $\hbar f_{0,n}(x,p)$ is given in Fig.~\ref{fig:fwign}. 

The classical limit can be implemented by taking $n\to+\infty$, $\hbar\to 0$ so that $\hbar  k_n(\eta,m_0,m_1)$ is kept fixed \cite{CaKi87, HaKo89, MMSSV05, MoPi21, MvdV21, CFL23}. Formally, it is obtained by setting
\begin{equation}\label{eq:hbar}
\hbar=\frac{ p_\text{c}}{ k_n(\eta,m_0,m_1)}\,,
\end{equation}
where $p_\text{c}$ is a reference value of the classical momentum, related to the classical reference energy  $E_\text{c}$ via 
\begin{equation}
E_\text{c}=\frac{p_\text{c}^2}{2m}\,,
\end{equation}
and by letting $n\to+\infty$.

By plugging Eq.~\eqref{eq:hbar} into Eq.~\eqref{eq:fn2}, we get for all $s \in \{-1,0,1\}$:
\begin{equation}
f_{s,n}(x,p)=\frac{\triangle(x) k_n(\eta,m_0,m_1)}{\pi  p_\text{c}}
\sinc\biggl(\frac{ k_n(\eta,m_0,m_1)}{p_\text{c}}(p-sp_\text{c})(1-2\abs{x})\biggr)\,, \quad \forall x,p \in \R.
\end{equation}
Then, by using the well-know  identity
\begin{align}
\lim_{a\to 0}\frac{1}{\pi}\frac{\sin(x/a)}{x} =\lim_{a\to 0}\frac{1}{\pi a}\sinc\Bigl(\frac{x}{a}\Bigr) =\ddelta(x)
\end{align}
to be understood in the distributional sense, we obtain that 
\begin{align}
\lim_{n\to+\infty} f_{s,n}(x,p)=\ind(x)\ddelta(p-s p_\text{c})\,,
\end{align}
with $\ddelta(x)$ being the Dirac delta distribution. Moreover, by the Riemann-Lebesgue lemma~\cite{Li58} and by~\eqref{eq:normC} one also gets
\begin{equation}
  \frac{C^{+}_{U,n}\overline{C^{-}_{U,n}}}{N^2_{U,n}}\, \e^{2\iu  k_n(\eta, m_0,m_1) x} \to 0
\end{equation}
in the distributional sense, as $n\to+\infty$. Therefore, as distributions,
\begin{align}
\lim_{n \to + \infty}W \psi_{U,n}(x,p)- \ind(x)\bigl[
 \omega_{U,n}  \ddelta(p-p_\text{c})  +(1-\omega_{U,n}) \ddelta(p+p_\text{c}) \bigr]=0 \,,
\end{align}
where we introduced the shorthand
\begin{equation}\label{eq:Omegan}
\omega_{U,n}= \frac{\abs{C_{U,n}^{+}}^2}{N_{U,n}^2}\,,
\end{equation}
and we used Eq.~\eqref{eq:normC}. Notice how, in the classical limit,  the information regarding the quantum BCs is all contained in the coefficients $\omega_{U,n}$. If it happens that the sequence $(\omega_{U,n})_{n\ge 1}$ admits a limit, say
\begin{equation}\label{eq:Ulim}
\lim_{n\to+\infty} \omega_{U,n}= \omega_U\in [0,1]\,,
\end{equation}
then we also get a well-defined distributional limit for the Wigner function, that is,
\begin{align}\label{eq:limWU}
W_U(x,p)\defeq \lim_{n\to+\infty} W\psi_{U,n}(x,p) 
= \ind(x)\bigl[\omega_U \ddelta(p-p_\text{c}) +(1-\omega_U) \ddelta(p+p_\text{c}) \bigr] \,.
\end{align}

As it turns out, the limit in Eq.~\eqref{eq:Ulim} does not exist for all the BCs $U$. To determine the classical limit of the Wigner function, hence, we have to finely examine the asymptotic behavior of the coefficients $\omega_{U,n}$ in the high-energy regime,  which in turn depends on the asymptotic behavior of the spectrum. We perform this analysis in Sec.~\ref{sec:asymptotic}. Before proceeding, however, in the next subsection we suggest  a classical interpretation of the limit Wigner function~\eqref{eq:limWU}. 

\subsection{Classical particle in a box}\label{sec:classical}
\begin{figure}[tbp]
\centering
\includegraphics{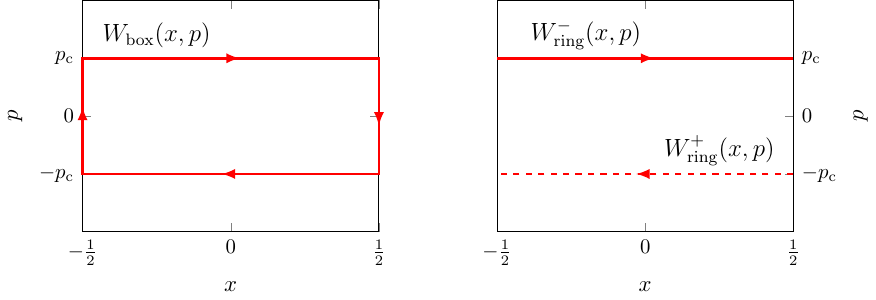}

\caption{Joint probability distribution of a classical particle in a box (left) and in a ring (right); in the ring, the dashed (solid) line represent a  (counter)clockwise motion.}
    \label{fig:classphase}
\end{figure}

Let us briefly review the phase-space picture in the classical setting~\cite{BeDoRo04}. Heuristically, the joint (stationary) probability distribution of a classical particle of mass $m$ and energy $E_\text{c}$ which is confined in a box of unit length with elastically reflecting hard walls
is given by 
\begin{subequations}\label{eq:Wbox}
\begin{align}
W_\text{box}(x,p)&=\ind(x)\sqrt{2mE_\text{c}}\ddelta(p^2-2mE_\text{c})\\
&=\frac{\ind(x)}{2}[\ddelta(p-p_\text{c})+\ddelta(p+p_\text{c})]\,,
\end{align}
\end{subequations}
where $p_\text{c}=\sqrt{2mE_\text{c}}$, and corresponds to a rectangular orbit in the phase-space, see the left panel of Fig.~\ref{fig:classphase}. If the particle is confined in a ring, instead, we can consider two classical orbits, associated with the joint probability distributions
\begin{equation}
W_\text{ring}^{\pm}(x,p)=\ind(x)\ddelta(p\mp p_\text{c}) \,,
\end{equation}
with $W_\text{ring}^{+}(x,p)$ and $W_\text{ring}^{-}(x,p)$ describing respectively a clockwise orbit and a counterclockwise one, see the right panel of Fig.~\ref{fig:classphase}. 

\begin{figure}[tbp]
\centering
\includegraphics{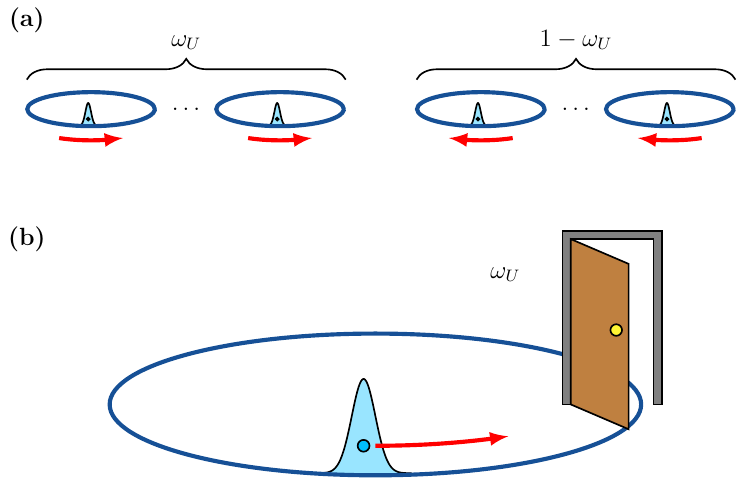}

\caption{Two possible classical realization of the limit Wigner function $W_U(x,p)$: (a) statistical ensemble of particles in a ring; (b) time-average of a single particle in a ring with a door (a classical junction) which opens with probability $\omega_U$.}
\label{fig:demon}
\end{figure}

Accordingly, the limit Wigner function $W_U(x,p)$ introduced in Eq.~\eqref{eq:limWU}, when it exists,  can be interpreted from a classical perspective in two different ways. Since
\begin{equation}\label{eq:WU}
W_U(x,p)= \omega_U W_\text{ring}^{+}(x,p)+(1-\omega_U) W_\text{ring}^{-}(x,p)\,,
\end{equation}
we can indeed consider $W_U(x,p)$ as the probability distribution of a classical ensemble of particles in a ring, of which a fraction $\omega_U$ is moving clockwise whereas the remaining fraction $1-\omega_U$ is moving counterclockwise, see Fig.~\ref{fig:demon} (a).

Another interesting interpretation is suggested by the ergodic theorem~\cite{Moo15}:  $W_U(x,p)$ can also represent the time-averaged probability distribution of a single classical particle in a ring with a junction, which acts as a door that can be opened or closed, allowing respectively the particle to pass through it or to be elastically reflected, thus inverting its motion, see Fig.~\ref{fig:demon} (b). In particular, in order to implement the limit distribution $W_U(x,p)$, each time the particle approaches the junction, the door has to be closed (and then subsequently reopened) with probability $\omega_U$ if the particle is moving clockwise, and with probability $1-\omega_U$ if it is moving counterclockwise. Notice however that, if the initial conditions are known, the probability distribution~\eqref{eq:WU} can be also realized by a \emph{deterministic} classical system.

\section{Asymptotic analysis}\label{sec:asymptotic} 
In order to analyze the asymptotic behavior of the coefficients $\omega_{U,n}$ for large $n$, we start by observing that Eq.~\eqref{eq:Cpm} implies that
\begin{equation} \label{eq:Cpmabs}
\abs{C^\pm_{U}(k)}=\abs{k A^{\pm}_U(k)\pm B^{\pm}_U(k) }\,,
\end{equation}
where
\begin{subequations} \label{eq:ApmBpm}
\begin{align}
A^{\pm}_U(k)&\defeq m_0+\iu m_3 -\e^{-\iu\eta}-\e^{\mp\iu k}(m_2+\iu m_1)\,,\\
B^{\pm}_U(k)&\defeq m_0+\iu m_3 +\e^{-\iu\eta}+\e^{\mp\iu k}(m_2+\iu m_1)\,.
\end{align}
\end{subequations}
Notice that $A^{\pm}_U(k)$ can vanish, as it happens for example when $m_0=\cos(\eta)$ and $m_3=-\sin(\eta)$ (and thus $m_1=m_2=0$). Therefore there are two possibilities for $\omega_{U,n}$, in the high-energy regime: 
\begin{subequations}\label{eq:omegalimits}
\begin{align} 
\lim_{n\to+\infty} \left(\omega_{U,n}-\frac{\abs{A^{+}_{U,n}}^2}{\abs{A^{+}_{U,n}}^2+\abs{A^{-}_{U,n}}^2}\right)=0\,,&&\text{if}\, A^\pm_{U,n}\neq0\,,\\
\lim_{n\to+\infty} \left(\omega_{U,n}-\frac{\abs{B^{+}_{U,n}}^2}{\abs{B^{+}_{U,n}}^2+\abs{B^{-}_{U,n}}^2} \right)=0\,,&&\text{if}\, A^\pm_{U,n}=0\,,
\end{align}
\end{subequations}
where $A_{U,n}^\pm\defeq A_{U}^\pm(k_n(\eta,m_0,m_1))$, $B_{U,n}^\pm\defeq B_{U}^\pm(k_n(\eta,m_0,m_1))$, and we used the fact that  $k_n(\eta,m_0,m_1)\to +\infty$ for $n\to +\infty$. 

We say that the sequence $(\omega_{U,n})_{n\ge1}$ is \emph{balanced}, when it admits the limit
\begin{equation}\label{eq:omegatriv}
\lim_{n\to+\infty} \omega_{U,n}=\frac{1}{2}\,,
\end{equation}
and is \emph{unbalanced} otherwise. 

The balanced case can be easily characterized: Eq.~\eqref{eq:omegatriv} holds when
\begin{subequations}\label{eq:ABlimits}
\begin{align} 
\lim_{n\to+\infty} \frac{\abs{A_{U,n}^-}^2}{\abs{A_{U,n}^+}^2}=1\,,&&\text{if}\, A^\pm_{U,n}\neq0\,,\\[2pt]
\lim_{n\to+\infty}  \frac{\abs{B_{U,n}^-}^2}{\abs{B_{U,n}^+}^2}=1\,,&&\text{if}\, A^\pm_{U,n}=0\,.
\end{align}
\end{subequations}
In turn, one can easily verify that these limits hold when one of the following sufficient  conditions is satisfied.
\begin{enumerate}
\item If $m_1=0$ and $m_2=0$, that is if we consider the asymmetric Robin BCs
\begin{align}\label{eq:Ulocal}
U\bigl(\eta, \cos(\beta),0,0,\beta\bigr)=\matt{\e^{\iu(\eta+\beta)}&0\\ 0& \e^{\iu(\eta-\beta)}}\,, 
\end{align}
then $A_{U,n}^+=A_{U,n}^-$ and $B_{U,n}^+=B_{U,n}^-$ for each $n\ge1$.
Remarkably, these are the most general local BCs. Notice that they can be either regular (if e.g\ $\beta=0$, when they reduce to symmetric Robin BCs) and singular (if e.g.\ $\beta=-\eta\neq 0$, which gives a mixed Dirichlet-Robin BC).
\item If 
\begin{equation} \label{eq:knlimit}
\lim_{n\to +\infty} \sin\bigl(k_n(\eta,m_0,m_1)\bigr
)=0
\end{equation}
then  $A_{U,n}^+\sim  A_{U,n}^-$ and $B_{U,n}^+\sim  B_{U,n}^-$ asymptotically as $n\to+\infty$. As we will show in the next subsection, this spectral condition is always satisfied for regular BCs.
\end{enumerate}
On the other hand, the study of the unbalanced case is more involved, and it requires the asymptotic estimate of the spectral quantities $\e^{\mp \iu k_n(\eta,m_0,m_1)}$ for large $n$.  Thus, we devote Sec.~\ref{sec:specasym} to analyze in detail this spectral asymptotics. Then, in Sec.~\ref{sec:wigasym}, after gathering the obtained results, we finally classify the possible classical limits of the Wigner functions $W_{U,n}(x,p)$.
\subsection{Spectral asymptotics}\label{sec:specasym}
\begin{figure}[tpb]
\centering
\includegraphics{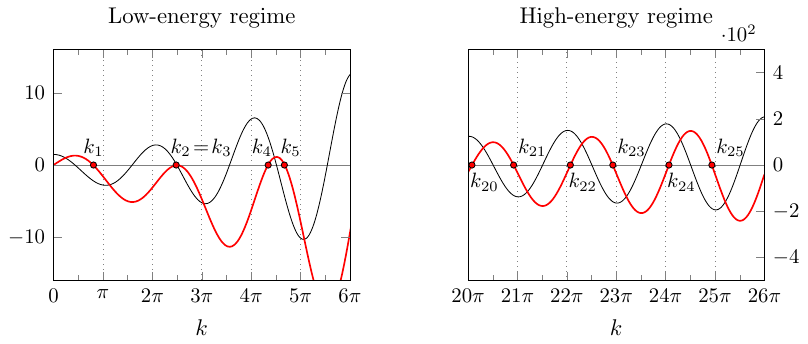}

\caption{Plot of the spectral function $F_U(\sqrt{k})$ in Eq.~\eqref{eq:FU} (red line) and of its derivative (black line) for the boundary condition $U(0,\cos \theta,\sin \theta,0)$ with $\theta\approx 0.25$.}
    \label{fig:exception}
\end{figure}

By defining the sequence
\begin{align}\label{eq:dn}
\delta_n(\eta, m_0,m_1)\defeq k_n(\eta,m_0,m_1)-n\pi,  \quad n \geq 1\,,
\end{align}
the spectral condition in Eq.~\eqref{eq:knlimit}, that is relevant for the balanced case, is equivalent to
\begin{equation} \label{eq:sinlimit}
\lim_{n\to +\infty} \sin\bigl(\delta_n(\eta,m_0,m_1)\bigr)=0\,,
\end{equation}
whereas the spectral quantities that are relevant for the unbalanced case can be rewritten as
\begin{align}\label{eq:ktodelta}
\e^{\mp \iu k_n(\eta,m_0,m_1)}=(-1)^n \e^{\mp \iu \delta_n(\eta,m_0,m_1)}\,.
\end{align}
As it turns out,  the behavior of $\bigl(\delta_n(\eta, m_0,m_1)\bigr)_{n \geq 1}$ can be quite erratic for small values of $n$, but becomes more regular for large $n$, see Fig.~\ref{fig:exception} for an example. The currently available spectral estimates associated with the Weyl law for quantum graphs \cite{BE09, OdSc19}  are not enough to characterize the remainder term in Eq.~\eqref{eq:dn}, as they just imply that $\delta_n(\eta,m_0,m_1)=o(n)$, for the system under study  (which can be regarded as the ``building block'' of more complex quantum graphs). However, in the particular situation considered here, the asymptotic behavior of $\delta_n$ is known, see Sec.~1.5 of~\cite{Mar86}. In this subsection we thus determine, on the lines of~\cite{Mar86}, the asymptotics of $\delta_n(\eta,m_0,m_1)$ needed to analyze the limit of the coefficients $\omega_{U,n}$.

To achieve this, let us rewrite the spectral function as
\begin{equation}
F_U(\epsilon)=a_U( k) k^2+b_U( k) k+c_U( k)\,, \quad \epsilon >0\,,
\end{equation}
where $ k = \sqrt{\epsilon}$ and
\begin{subequations}
\begin{align}
a_U( k)&\defeq \sin( k)\bigl(\cos(\eta)-m_0\bigr)\,,\\ 
b_U( k)&\defeq -2\bigl(m_1 -\sin(\eta)\cos( k)\bigr)\,,\\
c_U( k)&\defeq\sin( k)\bigl(\cos(\eta)+m_0\bigr)\,.
\end{align}
\end{subequations}
It is convenient to separate the analysis into three cases:
\begin{itemize}
\item BCs $U$ such that  $ k_n(\eta,m_0,m_1)= n \pi$ and hence $a_U\bigl( k_n(\eta,m_0,m_1)\bigr)=0$ for all $n \geq 1$ (exact cases including Dirichlet BCs);
\item BCs $U$ such that  $\cos\eta=m_0 \neq 1$ and hence $a_U( k)=0$ for all $ k >0$ (singular BCs);
\item all the remaining BCs $U$ (regular BCs).
\end{itemize}

\subsubsection{Exact cases.}
We first consider the case of BCs $U=U(\eta,m_0,m_1,\beta)$ such that $ k_n(\eta,m_0,m_1)=n\pi$ for all $n \geq 1$. We show that this case occurs if and only if $\eta=m_1=0$. In fact, \begin{equation}
F_U\bigl((2n\pi)^2\bigr)=-4n\pi[m_1-\sin(\eta)] =0 \quad\Leftrightarrow\quad m_1=\sin(\eta)\,,
\end{equation}
and \begin{equation}
F_U\bigl((2n+1)^2\pi^2\bigr)=-2 (2n+1)\pi[m_1+\sin(\eta)]=0 \quad\Leftrightarrow\quad m_1=-\sin(\eta)\,,
\end{equation}
and hence $m_1=\eta=0$. In this case we have that for all $n \geq 1$:
\begin{align}\label{eq:es1}
\delta_n(0,m_0,0)=0\,,
\end{align}
which clearly implies the spectral condition~\eqref{eq:knlimit}. Notice that the corresponding BCs $U(0,m_0,0,\beta)$ are always regular.

\subsubsection{Singular boundary conditions.}
Let us now consider the singular BCs~\eqref{eq:sing}, i.e.\ $\cos(\eta)=m_0 \neq 1$. For this choice of parameters the spectral function simplifies to
\begin{equation}
F_U(\epsilon)=2\sin ( k) \cos (\eta)-2 k\bigl[m_1-\sin (\eta) \cos  (k) \bigr]\,,
\end{equation}
and by considering the equation 
\begin{equation}
F_U\bigl( k_n(\eta,m_0,m_1)^2\bigr)=0\,,
\end{equation}
which, since $\eta\in(0,\pi)$, can be rearranged as
\begin{equation}
\cos\bigl(k_n(\eta, m_0, m_1)\bigr)-\frac{m_1}{\sin(\eta)}=-\frac{\cot(\eta)\sin\bigl(k_n(\eta, m_0, m_1)\bigr)}{ k_n(\eta, m_0, m_1)}\,,
\end{equation}
we get
\begin{equation}
\Bigl|\cos\bigl(k_n(\eta, m_0, m_1)\bigr)- \frac{m_1}{\sin (\eta)}\Bigr|\le \frac{\cot(\eta)}{ k_n(\eta, m_0, m_1)}\,.
\end{equation}
Notice that $\abs{m_1/\sin(\eta)}\le 1$. In this case, one can show \cite{Mar86} that for large $n$  the sequence $\bigl(k_{2n-1}(\eta, m_0, m_1), k_{2n}(\eta, m_0, m_1)\bigr)_{n\ge 1}$ is asymptotically close to the sequence 
 \begin{align}
\left(2n\pi - \arccos\biggl(\frac{m_1}{\sin(\eta)}\biggr),\, 2n\pi + \arccos\biggl(\frac{m_1}{\sin(\eta)}\biggr)\right)_{n\ge 1}\,.
\end{align}
In our notations, since $\arccos(x)=\pi-\arccos(-x)$, we can restrict $\delta_n(\eta, m_0, m_1)$ to $[0,\pi]$ obtaining the asymptotic limits
\begin{align}
\lim_{n \to +\infty} \delta_{2n}(\eta,\cos \eta,m_1)=\arccos\biggl(\frac{m_1}{\sin(\eta)}\biggr)\,,\label{eq:sbcasym}
\end{align}
and
\begin{align}
\lim_{n \to +\infty} \delta_{2n-1}(\eta,\cos \eta,m_1)= \arccos\biggl(-\frac{m_1}{\sin(\eta)}\biggr)\,.\label{eq:sbcasymd}
\end{align}
Remarkably, for $\eta=\pi/2$ we recover the exact spectral sequence \cite{BFV01}, that is:
\begin{equation}\label{eq:es2}
\delta_n\left(\frac{\pi}{2},0,m_1\right)=\arccos\bigl((-1)^{n}m_1\bigr)\,.
\end{equation}
For $m_1=0$, in particular, the correction is constant:
\begin{align}\label{eq:ppcond}
\delta_n\left(\frac{\pi}{2},0,0\right)=\frac{\pi}{2}\,.
\end{align}
Conversely, the limit $\eta\to 0$ (which gives Dirichlet BC) is ill-defined, and one should rely on the exact expression~\eqref{eq:es1}. 
We conclude that the asymptotic behavior of the sequence $\bigl(\delta_n(\eta,m_0,m_1)\bigr)_{n\ge1}$ for singular BCs  has a residual dependence on $U$, and, for $m_1\neq 0$, also on the parity $(-1)^n$ of $n$.

\subsubsection{Regular boundary conditions.}
For what concerns the remaining (regular) BCs,  we can now assume that $ k\neq n\pi$, as the latter values have been discussed before. We rewrite the equation $F_U(\epsilon)=0$ as
\begin{equation}
a_U(\epsilon)=\frac{b_U(\epsilon)}{ k}+\frac{c_U(\epsilon)}{ k^2}\,,
\end{equation}
from which we get
\begin{equation}
 \bigl\lvert\sin (k)\bigl(\cos(\eta)-m_0\bigr)\bigr\rvert=\left\lvert\frac{-2\bigl(m_1 -\sin(\eta)\cos (k)\bigr)}{ k}+\frac{\sin (k)\bigl(\cos(\eta)+m_0\bigr)}{ k^2}\right\rvert\le \left\lvert\frac{6}{k}\right\rvert\,.
\end{equation}
Therefore for all $n \geq 1$:
\begin{equation}
 \bigl\lvert\sin\bigl( k_n(\eta, m_0,m_1)\bigr)\bigr\rvert= \bigl\lvert\sin\bigl(\delta_n(\eta, m_0,m_1)\bigr) \bigr\rvert \leq \frac{6}{\abs{ \cos(\eta)-m_0}}\frac{1}{k_n(\eta,m_0,m_1)}\,.
\end{equation}
Then, by using the fact that the wave numbers accumulate to $+\infty$, the above inequality implies the spectral condition in Eq.~\eqref{eq:knlimit}, that is
\begin{align}\label{eq:regularlimit}
\lim_{n \to + \infty}\sin\bigl(\delta_n(\eta,m_0,m_1)\bigr)= 0.
\end{align}

\subsubsection{Numerical results.}
\begin{figure}[tpb]
\centering
\includegraphics{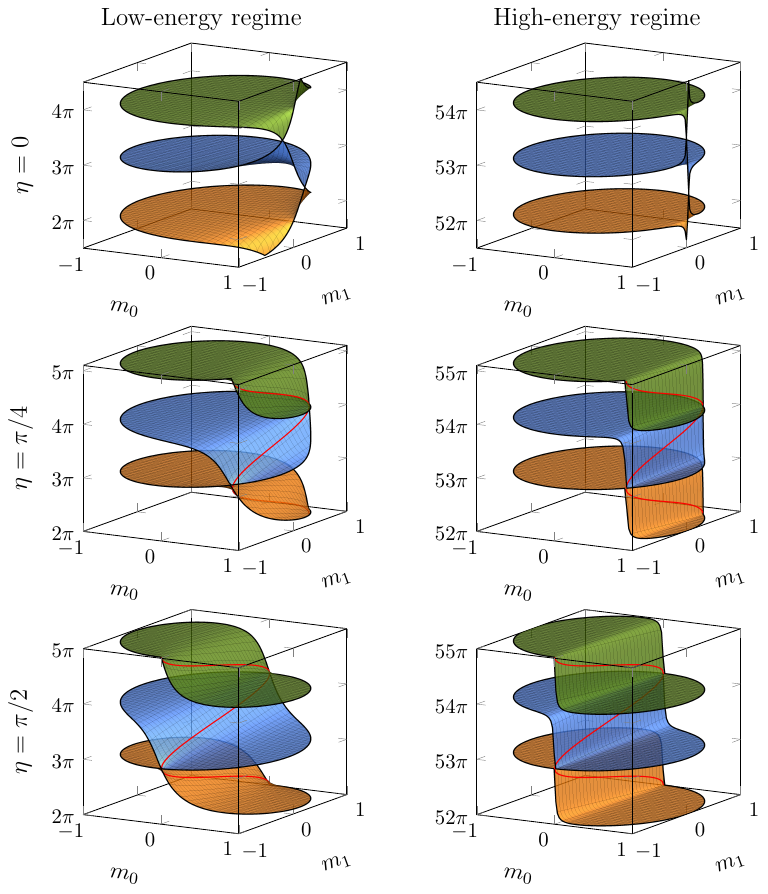}

\caption{Exact values of some $ k_n(\eta, m_0, m_1)$, plotted over the parameter space $(m_0,m_1)\in D$ for $\eta=0$ (top row), $\eta=\frac{\pi}{4}$ (middle row) and $\eta=\frac{\pi}{2}$ (bottom row). For $\eta\neq 0$, red lines have been added representing the asymptotic formulae~\eqref{eq:sbcasym}--\eqref{eq:sbcasymd}.}
    \label{fig:spectra}
\end{figure}

\begin{figure}[tpb]
 \centering
 \includegraphics{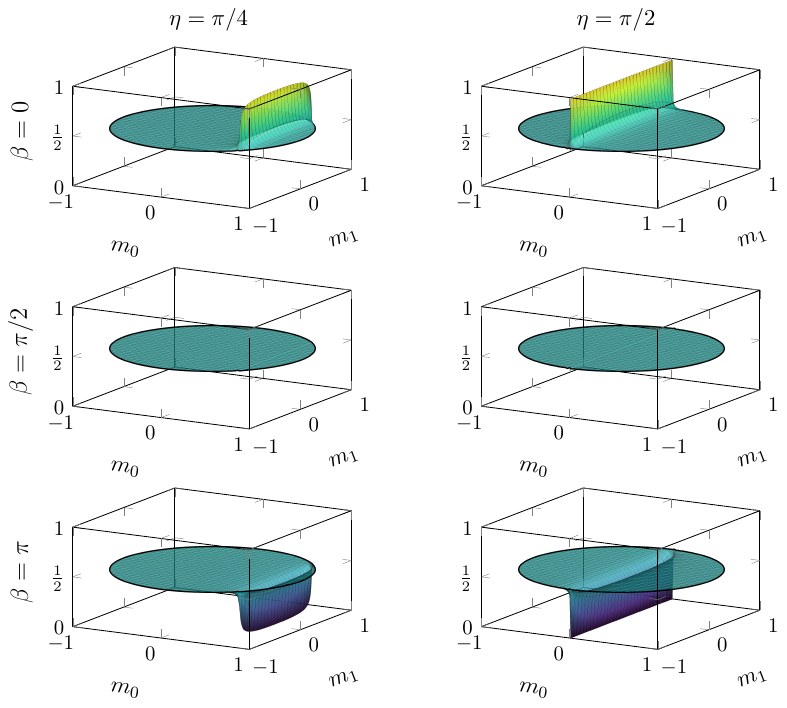}
 
 \caption{Exact values of $\omega_{U,52}$, plotted over the parameter space $(m_0,m_1)\in D$  for $\eta=\frac{\pi}{4}$ (left column) and $\eta=\frac{\pi}{2}$ (right column) and for different values of the non-spectral parameter $\beta$.}
     \label{fig:coeff}
 \end{figure}

To corroborate the asymptotic analysis, in Fig.~\ref{fig:spectra} we plot the values of some wave numbers $ k_n(\eta,m_0,m_1)$, which have been determined by numerically finding the zeroes of the spectral function, both in the low-energy regime (small $n$) and in the high-energy regime (large $n$), as function of the spectral parameters $\eta,m_0,m_1$. The high-energy plots are consistent with the asymptotic formulas obtained so far.
In Fig.~\ref{fig:coeff} we also plot some exact values of $\omega_{U,n}$ in the high-energy regime, again as function of the spectral parameters $\eta, m_0, m_1$. As expected, for $m_0\neq\cos(\eta)$  (that is for regular BCs) we have that $\omega_{U,n}\approx 1/2$, while for $m_0=\cos(\eta)$ we observe a residual dependence on $U$, and in particular on $\beta$.

\subsection{Asymptotics of the Wigner function} \label{sec:wigasym} 
\begin{figure}[tpb]
 \centering
 \includegraphics{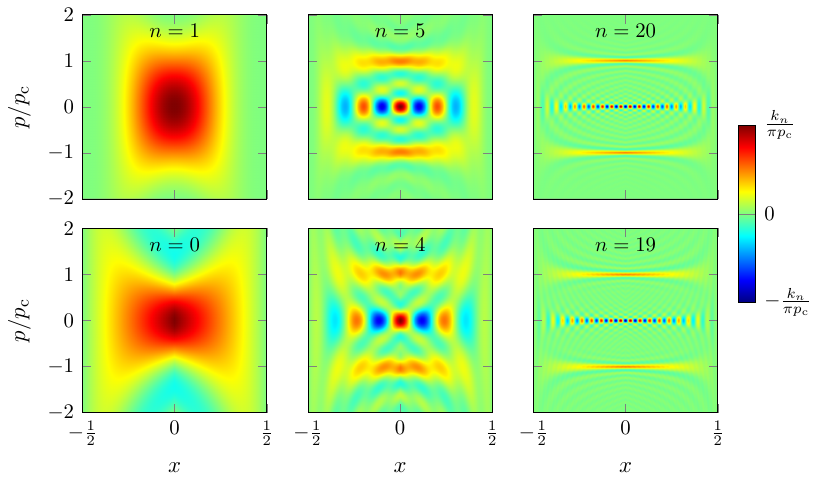}

 \caption{Density plot of $W\psi_{U,n}(x,p)$ for Dirichlet ($U=I$, first row) and Neumann ($U=-I$, second row) BCs; from left to right the value of $n$ is increased, by setting $\hbar=p_\text{c}/ k_n$, approaching the classical limit.}
     \label{fig:wReg}
 \end{figure}

To sum up, we obtained two sufficient conditions for having a balanced classical limit. Indeed, we found that if $U$ is a local BC, that is if $U=U\bigl(\eta, \cos(\beta),0,0,\beta\bigr)$ for  $\eta\in[0,\pi)$ and $\beta\in[0,2\pi)$, see Eq.~\eqref{eq:Ulocal}, or if $U$ is a regular BC, so that the spectral condition in Eq.~\eqref{eq:knlimit} is satisfied [see Eqs.~\eqref{eq:es1} and~\eqref{eq:regularlimit}], then the coefficients $\omega_{U,n}$ have the well-defined high-energy limit
\begin{equation}\label{eq:omeganreg}
\omega_U=\lim_{n\to+\infty}\omega_{U,n} =\frac{1}{2}\,,
\end{equation}
and the corresponding Wigner functions admit a limit in the form of Eq.~\eqref{eq:limWU} with balanced coefficients $\omega_U=1-\omega_U=1/2$, that is: 
\begin{equation}\label{eq:wlimreg}
\lim_{n\to+\infty} W\psi_{U,n}(x,p)=\frac{\ind(x)}{2}[\ddelta(p-p_\text{c})+\ddelta(p+p_\text{c})]\,. 
\end{equation}
This balanced classical limit is represented, respectively for the case of Dirichlet and Neumann BCs, in the first and in the second row of Fig.~\ref{fig:wReg}.

For singular \emph{non-local} BCs, instead, the situation is complicated by the fact that even in the high-energy regime the correction $\delta_n(\eta,m_0,m_1)$  does generally still depend on the parity of $n$, see Eqs.~\eqref{eq:sbcasym}--\eqref{eq:sbcasymd},  thus not admitting a limit. However, since in the high-energy regime also the coefficients $\omega_{U,n}$ depend on $n$ only through its parity, see Eqs.~\eqref{eq:Cpmabs}--\eqref{eq:ApmBpm} and~\eqref{eq:ktodelta}, the following limits
\begin{align}\label{eq:omeeo}
\omega_{U,\text{e}} \defeq \lim_{n\to+\infty} \omega_{U,2n}\,, && 
\omega_{U,\text{o}} \defeq \lim_{n\to+\infty} \omega_{U,2n+1}\,,
\end{align}
are well-defined and finite. We stress that both $\omega_{U,2n}$ and $\omega_{U,2n+1}$ have a limit, but in general $\omega_{U,2n}\neq \omega_{U,2n+1}$. Accordingly, although for singular non-local BCs the  Wigner functions $W\psi_{U,n}(x,p)$ do  not generally admit a classical limit,  the even and odd subsequences have the well-defined limits in the form of Eq.~\eqref{eq:limWU}, 
\begin{subequations}\label{eq:wlimsing}
\begin{align}
W_{U,\text{e}}(x,p)&\defeq \lim_{n\to+\infty} W\psi_{U,2n}(x,p)\notag\\
&
=\ind(x)\bigl[
\omega_{U,\text{e}} \ddelta(p-p_\text{c}) +(1- \omega_{U,\text{e}}) \ddelta(p+p_\text{c})  \bigr]\,,\\[5pt]
W_{U,\text{o}}(x,p)&\defeq\lim_{n\to+\infty} W\psi_{U,2n+1}(x,p)\notag\\
&
=\ind(x)\bigl[
  \omega_{U,\text{o}} \ddelta(p-p_\text{c}) +(1- \omega_{U,\text{o}}) \ddelta(p+p_\text{c}) \bigr]\,,
\end{align}
\end{subequations}
with (generally) unbalanced coefficients $\omega_{U,\text{e}}\neq 1/2$ and $\omega_{U,\text{o}} \neq 1/2$.

\begin{figure}[tpb]
 \centering
 \includegraphics{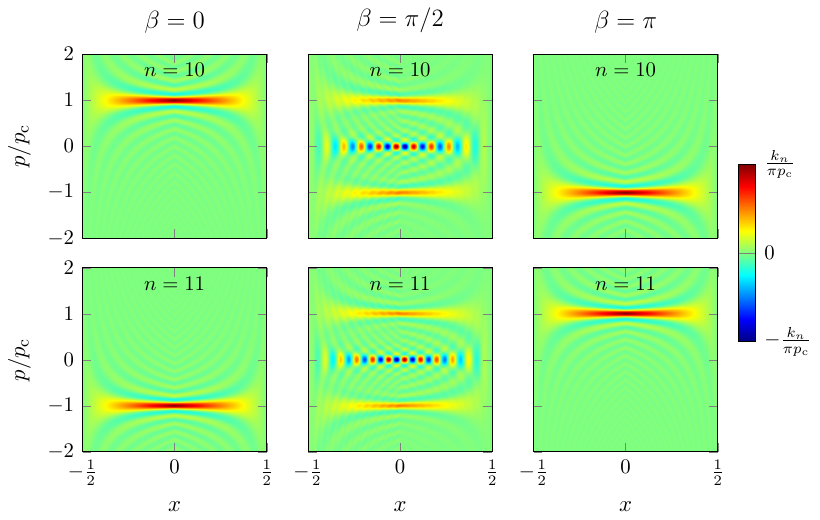}

 \caption{Density plot of $W\psi_{U,n}(x,p)$, with $\hbar=p_\text{c}/ k_n$, for the singular BCs $U(\tfrac{\pi}{2}, 0, 0,\beta)$, with $n=10$ (first row) and $n=11$ (second row) and for different values of the non-spectral parameter $\beta$.}
     \label{fig:wSing}
 \end{figure}

This phenomenon is shown in Fig.~\ref{fig:wSing} for the family of ``quasi-periodic'' BCs $U(\tfrac{\pi}{2},0,0, \beta)$, given by
\begin{align}
\psi\Bigl(\frac{1}{2}\Bigr)=\iu\cot\Bigl(\frac{\beta}{2}+\frac{\pi}{4}\Bigr)\psi\Bigl(-\frac{1}{2}\Bigr)\,,&&\psi'\Bigl(\frac{1}{2}\Bigr)=\iu\tan\Bigl(\frac{\beta}{2}+\frac{\pi}{4}\Bigr)\psi'\Bigl(-\frac{1}{2}\Bigr)\,.
\end{align}
In particular, these BCs reduce  for $\beta=0$ to the pseudo-periodic BC $U_\text{pp}(\pi/2)$, that is to $\psi(1/2)=\iu \psi(-1/2)$ and $\psi'(1/2)=\iu \psi'(-1/2)$, and to the mixed Dirichlet-Neumann BC $\psi(1/2)=0$ and $\psi'(-1/2)=0$ for $\beta=\pi/2$. Remarkably, for any $\beta\in[0,2\pi]$, by using Eq.~\eqref{eq:ppcond}  we are able to get the simple exact expression
 \begin{equation}\label{eq:ppcondome}
 \omega_{U(\frac{\pi}{2},0,0,\beta),n}=\begin{cases}
 \cos\bigl(\frac{\beta}{2}\bigr)^2\,,&n \text{ even}\\[5pt]
 \sin\bigl(\frac{\beta}{2}\bigr)^2\,,&n \text{ odd}
 \end{cases}\,,
 \end{equation}
corresponding to the limit Wigner functions
\begin{subequations}
\begin{align}
W_{U(\frac{\pi}{2},0,0,\beta),\text{e}}(x,p)&=\ind(x)\Bigl[\cos\Bigl(\frac{\beta}{2}\Bigr)^2\ddelta(p-p_\text{c})+\sin\Bigl(\frac{\beta}{2}\Bigr)^2\ddelta(p+p_\text{c})\Bigr]\,,\\
W_{U(\frac{\pi}{2},0,0,\beta),\text{o}}(x,p)&=\ind(x)\Bigl[\sin\Bigl(\frac{\beta}{2}\Bigr)^2\ddelta(p-p_\text{c})+\cos\Bigl(\frac{\beta}{2}\Bigr)^2\ddelta(p+p_\text{c})\Bigr]\,.
\end{align}
\end{subequations}

\section{Discussion and outlook}\label{sec:discussion}

We showed that in the classical limit both local boundary conditions and regular boundary conditions  are associated with a ``balanced'' ensemble, having a  limit coefficient $\omega_U=1/2$, so that the corresponding limit Wigner function coincides with the joint probability distribution of a classical particle in a box with elastically reflecting walls, \begin{equation}
\lim_{n\to\infty} W\psi_{U,n}(x,p)=W_\text{box}(x,p)\,.
\end{equation}
Notice that in the limit any information about the quantum boundary condition is lost. 
 In this sense, the whole family of quantum systems with local and regular boundary conditions correspond to one and the same classical system. 

For what concerns singular non-local boundary conditions, the situation is more elaborate, as the coefficients $\omega_{U,n}$ do not generally admit a limit, but oscillate between the limits of the the even and odd subsequences, that is between the values $\omega_{U,\text{e}}$ and $\omega_{U,\text{o}}$ defined in Eq.~\eqref{eq:omeeo}. The corresponding Wigner functions behave accordingly, 
with the even and odd subsequences having the limit Wigner functions $W_{U,\text{e}}(x,p)$ and $W_{U,\text{o}}(x,p)$  given by Eqs.~\eqref{eq:wlimsing}. These latter distributions have exactly the form~\eqref{eq:WU}, in general with an unbalanced coefficient $\omega_U\neq 1/2$ carrying a residual information---a classical echo---of the quantum boundary condition $U$. As we discussed in Sec.~\ref{sec:classical}, in this case, for a given parity, the limit Wigner function can be interpreted as the probability distribution of an ensemble of classical particles in a ring, with a fraction $\omega_U$ moving clockwise and a fraction $1-\omega_U$ moving counterclockwise.

We conclude with some outlooks. In this Article we have analyzed the classical limit for the eigenfunctions of the non-relativistic kinetic-energy operator in a one-dimensional box with general self-adjoint boundary conditions. The corresponding classical distribution probabilities are stationary, i.e.\ time-independent. Performing a similar analysis by considering suitable \emph{wave packets}, instead of the eigenfunctions, we expect to obtain a different classical distribution \cite{LeAlSc88, BeDoRo04} mimicking a classical dynamical orbit. Future research will be devoted to this subject. Besides, more generally, it is still an open question if in the classical limit different self-adjoint extensions of a given operator all collapse (in a suitable sense) to the same classical object. One could investigate this problem by looking at the asymptotic behavior of the symbols associated with the different self-adjoint extensions of the operator \cite{CFL23, CLS24}. Other interesting generalizations of the present work may involve the analysis of a particle with spin and of a relativistic particle in a box (with general boundary conditions) \cite{ShBiRa92, KoRe16, isodirac}, as well as the case of multiple particles  \cite{DeDeLe21}.

\section*{Acknowledgments}

This work was partially supported by Istituto Nazionale di Fisica Nucleare (INFN) through the project ``QUANTUM'', by PNRR MUR Project PE0000023-NQSTI, by Regione Puglia and QuantERA ERA-NET Cofund in Quantum Technologies (Grant No. 731473), project PACE-IN, by the Italian National Group of Mathematical Physics (GNFM-INdAM), and by the Italian funding within the ``Budget MUR - Dipartimenti di Eccellenza 2023--2027'' - Quantum Sensing and Modelling for One-Health (QuaSiModO).


\section*{References}
\begin{thebibliography}{99}
\bibitem{Wi32} E. Wigner, \emph{On the quantum correction for thermodynamic equilibrium}, Phys. Rev. \textbf{40}, 749--759 (1932).

\bibitem{HiCoSc84} M. Hillery, R.F. O'Connell, M.O. Scully  and E.P. Wigner, \emph{Distribution functions in physics: Fundamentals}, Phys. Rep. \textbf{106}, 121--167 (1984).

\bibitem{Lee95} H.-W. Lee, \emph{Theory and application of the quantum phase-space distribution functions}, Phys. Rep.  \textbf{259},  3 (1995).

\bibitem{CaKi87} G.G. Cabrera and M. Kiwi, \emph{Large quantum-number states and the correspondence principle}, Phys. Rev. A \textbf{36}, 6 (1987).

\bibitem{HaKo89} G.Q. Hassoun and D.H. Kobe, \emph{Synthesis of the Planck and Bohr formulations of the correspondence principle}, Am.  J. Phys. \textbf{57}, 658--662 (1989).

\bibitem{MMSSV05} V.I. Man'ko, G. Marmo, A. Simoni, A. Stern and  F. Ventriglia, \emph{Tomograms in the quantum-classical transition}, Phys. Lett. A \textbf{343}, 4 251--266 (2005).

\bibitem{KoZeGl20} B. Koczor, R. Zeier and S.J. Glaser, \emph{Continuous phase-space representations for finite-dimensional quantum states and their tomography}, Phys. Rev. A \textbf{101}, 022318 (2020).

\bibitem{PrBrTo14} M. Przanowski, P.  Brzykcy and J. Tosiek, \emph{From the Weyl quantization of a particle on the circle to number-phase Wigner functions}, Ann. Phys. \textbf{351},  919--934 (2014).

\bibitem{KoLa21} K. Kowalski and K. \L{}awniczak, \emph{Wigner functions and coherent states for the quantum mechanics on a circle}, J. Phys. A: Math. Theor. \textbf{54}, 275302 (2021).

\bibitem{ZhVo03} S. Zhang and A. Vourdas, \emph{Phase space methods for particles on a circle}, J. Math. Phys. \textbf{44}, 5084 (2003).

\bibitem{RSSK11} I. Rigas, L.L. S\'{a}nchez-Soto, A.B. Klimov, J. \v{R}eh\'{a}\v{c}ek, Z. Hradil, \emph{Orbital angular momentum in phase space},  Ann. Phys.  \textbf{326},  2 (2011).

\bibitem{Li16} M. Ligab\`{o}, \emph{Torus as phase space: Weyl quantization, dequantization, and Wigner formalism}, J. Math. Phys. \textbf{57}, 082110 (2016).

\bibitem{DiPr02} N.C. Dias  J.N. Prata \emph{Wigner functions with boundaries}, J. Math. Phys. \textbf{43}, 4602 (2002).

\bibitem{KrWa05} S. Kryukov and M.A. Walton, \emph{On infinite walls in deformation quantization}, Ann. Phys. \textbf{317},  474--491 (2005).

\bibitem{DiPr07} N.C. Dias and J.N. Prata, \emph{Deformation quantization of confined systems}, Int. J. Quantum Inf. \textbf{05}, 257--263 (2007).

\bibitem{DiPoPr11} N.C. Dias, A. Posilicano and J.N. Prata, \emph{Self-adjoint, globally defined Hamiltonian operators for  systems with boundaries}, Comm. Pure Appl. Math. \textbf{10}, 6 (2011).

\bibitem{CaKrMa91} M. Casas, H. Krivine and J. Martorell, \emph{On the Wigner transforms of some simple systems and their semiclassical interpretation}, Eur. J. Phys. \textbf{12}, 105 (1991).

\bibitem{BeDoRo04} M. Belloni, M.A. Doncheski and R.W. Robinett, \emph{Wigner quasi-probability distribution for the infinite square well: Energy eigenstates and time-dependent wave packets}, Am. J. Phys. \textbf{72}, 1183--92 (2004).
 
\bibitem{Wal07} M.A. Walton, \emph{Wigner functions, contact interactions, and matching}, Ann. Phys.  \textbf{322}, 9 (2007).

\bibitem{AlWi21} M.H. Al-Hashimi and U.-J. Wiese, \emph{Canonical quantization on the half-line and in an interval based upon an alternative concept for the momentum in a space with boundaries}, Phys. Rev. Research \textbf{3}, 033079 (2021).

\bibitem{Rob95} R.W. Robinett, \emph{Quantum and classical probability distributions for position and momentum}, Am. J. Phys. \textbf{63}, 9 (1995).

\bibitem{Rob02} R.W. Robinett, \emph{Visualizing classical and quantum probability densities for momentum using variations on familiar one-dimensional potentials}, Eur. J. Phys. \textbf{23}, 165 (2002).

\bibitem{BeMaGa13} J. Bernal, A. Mart\'{i}n-Ruiz and J. Garc\'{i}a-Melgarejo, \emph{A Simple Mathematical Formulation of the Correspondence Principle}, J. Mod. Phys. \textbf{4}, 1 (2013).

\bibitem{Te14} G. Teschl, \emph{Mathematical methods in quantum mechanics}, Graduate Studies in Mathematics, American Mathematical Society (2014).

\bibitem{BFV01}  G. Bonneau, J. Faraut and G. Valent, \emph{Self-adjoint extensions of operators and the teaching of quantum mechanics}, Am. J. Phys. \textbf{69}, 3 (2001).

\bibitem{AIM05} M. Asorey, A. Ibort and G. Marmo, \emph{Global theory of quantum boundary conditions and topology change}, Int. J. Mod. Phys. A \textbf{20}, 1001--1025 (2005).
 
\bibitem{AIM15} M. Asorey, A. Ibort and G. Marmo, \emph{The topology and geometry of self-adjoint and elliptic boundary conditions for Dirac and Laplace operators}, Int. J. Geom. Methods
Mod. Phys. \textbf{12}, 06 (2015).

\bibitem{IBP15} A. Ibort, F. Lled\'o and J.M. P\'erez-Pardo, Self-adjoint extensions of the Laplace-Beltrami operator and unitaries at the boundary, \emph{J. Funct. Anal.} \textbf{268}, 3 634--670 (2015). 

\bibitem{FGL18b} P. Facchi, G. Garnero and M. Ligab\`{o}, \emph{Self-adjoint extensions and unitary operators on the boundary}, Lett. Math. Phys. \textbf{108},  195--212 (2018).

\bibitem{Cas08} W.B. Case, \emph{Wigner functions and Weyl transforms for pedestrians}, Am. J. Phys. \textbf{76}, 937--946 (2008).

\bibitem{FGL18} P. Facchi, G. Garnero and M. Ligab\`{o}, \emph{Quantum cavities with alternating boundary conditions}, J. Phys. A: Math. Theor. \textbf{51}, 105301 (2018).

\bibitem{BW10}  B. Belchev and M.A. Walton, \emph{On Robin boundary conditions and the Morse potential in quantum mechanics}, J. Phys. A: Math. Theor., \textbf{43}, 8 (2010). 

\bibitem{BaBiMa95} A. Balachandran, G. Bimonte, G. Marmo, and A. Simoni, \emph{Topology change and quantum physics}, Nucl. Phys. B \textbf{446}, 299--314 (1995).

\bibitem{ShWi12} A.D. Shapere, F. Wilczek, and Z. Xiong, \emph{Models of Topology Change}, arXiv: 1210.3545 [hep-th] (2012).

\bibitem{AFMP13} M. Asorey, P. Facchi, G. Marmo, and S. Pascazio, \emph{A dynamical composition law for boundary conditions}, J. Phys. A: Math. Theor. \textbf{46}, 102001 (2013).

\bibitem{IbLlPe15} A. Ibort, F. Lled\'{o}, and J.M. P\'{e}rez-Pardo, \emph{Self-adjoint extensions of the Laplace-Beltrami operator and unitaries at the boundary}, J. Funct. Anal. \textbf{268}, 634--670 (2015).

\bibitem{FGMSS18} P. Facchi, G. Garnero, G. Marmo, J. Samuel, and S. Sinha, \emph{Boundaries without boundaries}, Ann. Phys. \textbf{394}, 139--154 (2018).


\bibitem{movingwalls} S. Di Martino, F. Anz\`{a}, P. Facchi, A. Kossakowski, G. Marmo, A. Messina, B. Militello and S. Pascazio, \emph{A quantum particle in a box with moving walls},
J. Phys. A: Math. Theor. \textbf{46}, 365301 (2013).

\bibitem{bangalectures} S. Di Martino and P. Facchi, \emph{Quantum systems with time-dependent boundaries}, Int. J. Geom. Methods Mod. Phys. \textbf{12}, 1560003 (2015).



\bibitem{isoboundary} G. Angelone, P. Facchi and G. Marmo, \emph{Hearing the shape of a quantum boundary conditions},  Mod. Phys. Lett. A \textbf{37}, 2250114 (2022).
 
 
 \bibitem{Ze05} A. Zettl, \emph{Sturm-Liouville Theory}, Mathematical surveys and monographs, American Mathematical Society (2005).

\bibitem{MoPi21} J. Mostowski and J. Pietraszewicz,  \emph{Wigner Function for Harmonic Oscillator and The Classical Limit}, arXiv:2104.06638 [quant-ph] (2021).

\bibitem{MvdV21} V. Moretti and C.J.F. van den Ven, \emph{The classical limit of Schr\"{o}dinger operators in the framework of Berezin quantization and spontaneous symmetry braking as an emergent phenomenon}, Int. J. Geom. Methods Mod. Phys. \textbf{19}, 01 (2022).

\bibitem{CFL23} F.D. Cunden, P. Facchi and M. Ligab\`{o}, \emph{The semiclassical limit of a quantum Zeno dynamics}, Lett. Math. Phys. \textbf{113}, 114 (2023).

\bibitem{CLS24} F.D. Cunden, M. Ligab\`{o} and M. C. Susca, \emph{Truncated quantum observables and their semiclassical limit}, in preparation (2024).

\bibitem{Li58} M.J. Lighthill, \emph{An Introduction to Fourier Analysis and Generalised Functions}, Cambridge University Press (1958).


\bibitem{Moo15} C.C. Moore, \emph{Ergodic theorem, ergodic theory, and statistical mechanics}, Proceedings of the National Academy of Sciences \textbf{112},  7 1907--1911 (2015). 

\bibitem{BE09} J. Bolte, S. Endres, \emph{The Trace Formula for Quantum Graphs with General Self Adjoint Boundary Conditions}, Ann. Henri Poincar\'{e} \textbf{10}, 189--223, (2009).

\bibitem{OdSc19} A. Od\v{z}ak, L. \v{S}\'{c}eta, \emph{On the Weyl Law for Quantum Graphs}, Bull. Malays. Math. Sci. Soc. \textbf{42}, 119--131 (2019). 

\bibitem{Mar86} V.A. Marchenko, \emph{Sturm-Liouville Operators and Applications}, Birkh\"auser Basel (1986). 

\bibitem{LeAlSc88} C. Leubner, M. Alber and N. Schupfer, \emph{Critique and correction of the textbook comparison between classical and quantum harmonic oscillator probability densities}, Am. J. Phys. \textbf{56}, 1123--1129 (1988).

\bibitem{ShBiRa92}  G.R. Shin, I. Bialynicki-Birula and J. Rafelski, \emph{Wigner function of relativistic spin-1/2 particles},  Phys. Rev. A \textbf{46}, 645 (1992).


\bibitem{KoRe16} K. Kowalski and J. Rembieli\'{n}ski, \emph{The Wigner function in the relativistic quantum mechanics}, Ann. Phys. \textbf{375}, 1--15 (2016).



\bibitem{isodirac} G. Angelone, \emph{Hearing the boundary conditions of the one-dimensional Dirac operator}, arXiv:2311.17561 [quant-ph] (2023).

\bibitem{DeDeLe21} B. De Bruyne,  D.S. Dean, P. Le Doussal, S.N. Majumdar and G. Schehr, \emph{Wigner function for noninteracting fermions in hard-wall potentials}, Phys. Rev A \textbf{104}, 013314 (2021).


\end{thebibliography}
\end{document}